\documentclass[11pt]{article}

\global\arraycolsep=2pt

\usepackage[a4paper,top=30.6mm,bottom=38.6mm,left=34.6mm,right=34.6mm,footskip=1.3cm]{geometry}

\usepackage{mathrsfs,amsbsy,amssymb,latexsym,amsfonts,amsmath,cite}
\usepackage{graphicx,color}
\usepackage{mathtools}
\numberwithin{equation}{section}

\usepackage[inline]{enumitem}
\setitemize[1]{leftmargin=1em}
\usepackage{booktabs}

\allowdisplaybreaks

\newcommand{\ga}{\gamma}

\newcommand{\ep}{\epsilon}

\newcommand{\Tr}{{\rm Tr}}

\newcommand{\alg}[1]{\mathfrak{#1}}
\newcommand{\el}{\nonumber}
\newcommand{\nln}{\nonumber\\}

\usepackage{hyperref}

\usepackage{fancyhdr}
\newcommand{\preprint}[1]{%
  \fancypagestyle{empty}{%
    \fancyhf{} 
    \fancyheadoffset[R]{\marginparsep+\marginparwidth}
    \fancyhead[R]{{#1}}
    \renewcommand{\headrulewidth}{0pt}
    \renewcommand{\footrulewidth}{0pt}
  }
}

\newcommand*{\AdS}[1]{\ensuremath{\text{AdS}_{#1}}}
\newcommand{\dS}[1]{\ensuremath{\text{dS}_{#1}}}

\begin{document}

\preprint{KUNS-2559}


\begin{center}
{\Huge \textbf{Yang-Baxter deformations\\[.3em] of Minkowski spacetime} }
\vspace*{1.5cm}\\
{\large 
Takuya Matsumoto$^{\dagger}$, 
Domenico Orlando$^\sharp$, 
Susanne Reffert$^{\flat}$, 
\\ 
Jun-ichi Sakamoto$^{\ast}$
and Kentaroh Yoshida$^{\ast}$
} 
\end{center}
\vspace*{0.2cm}
\begin{center}
$^{\dagger}${\it 
Graduate School of Mathematics and Institute for Advanced Research, \\  
Nagoya University, Nagoya 464-8602, Japan}
\vspace*{0.25cm}\\ 
$^{\sharp}${\it IPT Ph.Meyer, and LPTENS\\
24 rue Lhomond, 75005 Paris, France}
\vspace*{0.25cm}\\ 
$^{\flat}${\it Institute for Theoretical Physics, 
Albert Einstein Center for Fundamental Physics, 
University of Bern, Sidlerstrasse 5, CH-3012 Bern, 
Switzerland 
}
\vspace*{0.25cm}\\ 
$^{\ast}${\it Department of Physics, Kyoto University, \\ 
Kyoto 606-8502, Japan} 
\end{center}
\vspace{0.3cm}

\begin{abstract}
We study Yang--Baxter deformations of 4D Minkowski spacetime. 
The Yang--Baxter sigma model description was originally developed for principal chiral models 
based on a modified classical Yang--Baxter equation. It has been extended to coset curved spaces 
and models based on the usual classical Yang--Baxter equation. 
On the other hand, for flat space, there is the obvious problem that the standard bilinear form degenerates 
if we employ the familiar coset Poincar\'e group/Lorentz group. 
Instead we consider a slice of \AdS5 
by embedding the 4D Poincar\'e group into the 4D conformal group $SO(2,4)$\,.   
With this procedure we obtain 
metrics and $B$--fields as Yang--Baxter deformations which correspond to 
well-known configurations such as T--duals of Melvin backgrounds,  
Hashimoto--Sethi and Spradlin--Taka\-yanagi--Volovich backgrounds, the T--dual of Grant space, 
pp-waves, and T--duals of \dS4 and \AdS4. Finally we consider a deformation 
with a classical $r$-matrix of Drinfeld--Jimbo type and explicitly derive 
the associated metric and $B$--field which we conjecture to correspond to a new integrable system.
\end{abstract}

\setcounter{footnote}{0}
\setcounter{page}{0}
\thispagestyle{empty}

\newpage
\tableofcontents
\newpage
\section{Introduction}

The integrability of 2D non-linear sigma models 
such as $O(N)$-invariant relativistic field theory~\cite{Luscher} 
has been extensively studied in various contexts (for a comprehensive book, see~\cite{AAR}). 
A prominent example is the AdS/\textsc{cft} correspondence~\cite{M}. The Green--Schwarz action of 
type IIB string theory on $\AdS5 \times S^5$, which is often abbreviated 
as the $\AdS5\times S^5$ superstring, was constructed in~\cite{MT} and has been shown to be classically integrable
in the sense that the Lax pair exists (\emph{i.e.} kinematical integrability)~\cite{BPR}. 
For related topics, see the review~\cite{review}.  

\medskip 

The study of integrable deformations of integrable 
non-linear sigma models is an interesting topic. Deformations of $S^3$ and \AdS3 have been well 
investigated~\cite{Cherednik,FR,BFP,Klimcik,KY,KYhybrid,
KMY-QAA,DMV,ORU,BR,KY-Sch,Kame,KOY,DMV-WZW,S-lambda,MY-YBE}.    
A systematic way is the Yang--Baxter sigma model approach proposed by Klimcik~\cite{Klimcik}. 
It was originally invented for principal chiral models based on the modified classical Yang--Baxter 
equation (m\textsc{cybe}). It was generalized to coset spaces~\cite{DMV} and then extended to the standard 
classical Yang--Baxter equation (\textsc{cybe}). In particular, squashed S$^3$~\cite{Cherednik} 
and 3D Schr\"odinger spacetime~\cite{Sch3} are associated with the m\textsc{cybe} and the \textsc{cybe}, 
respectively. 

\medskip 

An important application of the Yang--Baxter sigma model description are  
integrable deformations of the $\AdS5\times S^5$ superstring. 
A $q$-deformation with a classical $r$-matrix satisfying the m\textsc{cybe}  
has been studied in~\cite{DMV2}. Jordanian deformations based on the \textsc{cybe} have been 
proposed in~\cite{KMY-Jordanian-typeIIB}. In the latter case, there are a lot of 
classical $r$-matrices satisfying the \textsc{cybe} and some of them are associated with 
well-known gravitational backgrounds, such as Lunin--Maldacena--Frolov backgrounds~\cite{LM,Frolov}, 
gravity duals for non-commutative gauge theories~\cite{HI,MR}, 
Schr\"odinger spacetimes~\cite{MMT,BK,BKP} and gravity duals 
for dipole theories~\cite{dipole1,dipole2,dipole3,dipole4,Imeroni}, 
as shown in a series of works~\cite{LM-MY,MR-MY,Sch-MY,SUGRA-KMY,MY-duality}. 
Very recently, the reality of the classical action has been revisited in~\cite{Stijn} 
and a unified picture of deformed integrable sigma models has been provided in~\cite{Benoit}. 
It should be remarked that the Yang--Baxter sigma model approach works well 
even for non-integrable deformations. The case of $\AdS5 \times T^{1,1}$ 
is known to be non-integrable~\cite{BZ}. However, deformations of this background~\cite{LM,CO} 
can be reproduced as Yang--Baxter deformations~\cite{CMY}. 
The above relations between gravity solutions and classical $r$-matrices may be 
called the gravity/\textsc{cybe} correspondence (for a brief summary, see~\cite{MY-summary}).

\medskip 

It would be very interesting to generalize a correspondence of this type to the case of flat space, 
instead of $\AdS5 \times S^5$\,. In particular, one expects a relationship 
between certain classical $r$-matrices and Melvin twists~\cite{Melvin}.  
Therefore, in this paper, we consider Yang--Baxter deformations of 4D Minkowski spacetime. 
For flat space, we however immediately encounter a problem if we employ the familiar coset Poincar\'e group/Lorentz group, namely the degeneracy of the bilinear form. 
A possible resolution\footnote{Another way is to employ a generalized symmetric two-form~\cite{NW}. 
For a generalization to Schr\"odinger spacetimes, see~\cite{SYY}.}  is to consider instead a slice of \AdS5 
by embedding the 4D Poincar\'e group into the 4D conformal group $SO(2,4)$\,.
This embedding appears to work well, as we can reproduce the deformed metric 
and $B$--field associated to well-known backgrounds 
such as T--duals of Melvin backgrounds~\cite{Gibbons:1987ps,Tseytlin:1994ei,HT}, Hashimoto--Sethi backgrounds~\cite{HS}, 
time-dependent backgrounds of Spradlin--Takayanagi--Volovich~\cite{STV}, 
the T--dual of Grant space~\cite{Tai}, pp-wave backgrounds, 
and T--duals of \dS4 and $\AdS4$\,, as Yang--Baxter deformations. 
Finally we consider a deformation based on a classical $r$-matrix 
of Drinfeld--Jimbo type~\cite{Drinfeld1,Drinfeld2,Jimbo} 
and explicitly derive the associated metric and $B$--field.  

\medskip

In the following, we embed most of these 4D $\sigma$--models into string theory. 
In doing so, one can hope to understand the integrability of said models from a string theory perspective. 
We start with the simplest ansatz, in which we assume the extra six dimensions to be flat 
(as resulting from a $T^6$--compactification). Here, we will be looking for solutions 
which contain only a dilaton and no Ramond--Ramond--fields. It turns out that for a large class of the models studied, 
such a string theory embedding can be found. They all turn out to be $TsT$ transformations of flat space. 
As we are starting from an integrable background (\emph{i.e.} flat space with identifications) 
and T--duality preserves integrability~\cite{Ricci:2007eq, Mohammedi:2008vd, 
Curtright:1994be, Beisert:2008iq, Orlando:2012hu}, this fact explains the integrability of these models 
from a string theory point of view. 

\medskip 

For other classes of $r$-matrices, this is not the case, but they can instead be related to known integrable models with \dS4 or \AdS4 target space via T--duality. 
In order to complete them to full string theory solutions, the introduction of RR--fields is necessary 
(in the case of \dS4, even imaginary RR--form fields are needed), 
so these models cannot be captured with the simplest possible ansatz. 
A similar approach in the case of the $\AdS2 \times S^2$ geometry has been discussed in~\cite{LRT}, 
where type II theories are reduced to 4D theories plus a $T^6$\,. 


\medskip 

The plan of this paper is as follows. 
In Section~\ref{sec:def-4d-minkowski}, we introduce Yang--Baxter deformations of 4D Minkowski spacetime. 
In Sections ~\ref{sec:TsT-flat-space}-\ref{sec:Drinfeld-Jimbo}, we will provide some examples 
of classical $r$-matrices and the associated metrics and two-form $B$--fields. 
There is in particular the class of models that correspond to $TsT$--transformations of flat space 
(Sec.~\ref{sec:TsT-flat-space}). In Section~\ref{sec:dS-AdS}, we consider non-twist cases including the
S--dual of the pp-wave background, T--duals of \dS4 and \AdS4, and two-parameter deformations. 
In Section~\ref{sec:Drinfeld-Jimbo}, we extend the formulation from the \textsc{cybe} to the m\textsc{cybe} 
and then study a deformation with a classical $r$-matrix of Drinfeld--Jimbo type.  
Section~\ref{sec:conclusions} is devoted to conclusion and discussion. 

\section{Deformations of 4D Minkowski spacetime}\label{sec:def-4d-minkowski}

In this section we consider Yang--Baxter deformations of 2D sigma-model actions 
whose target space is given by 4D Minkowski spacetime. 

\subsection{Coset construction of \AdS5 revisited}\label{sec:coset-rev}

Let us remind ourselves of the coset construction of \AdS5 
with Poincar\'e coordinates. 

\medskip

It is well known that \AdS5 can be represented by a symmetric coset as
\begin{equation}
  \AdS5 = \frac{SO(2,4)}{SO(1,4)}
\end{equation}
and its metric can be computed via a coset construction.
To express the generators of the Lie algebras $\alg{so}(2,4)$ and $\alg{so}(1,4)$
it is necessary to first introduce some quantities. 

\medskip

We first introduce the gamma matrices $\gamma_{\mu}$ and $\gamma_5$ 
defined as  
\begin{align}
&\ga_1=\begin{pmatrix}
0&~0~&~0~&-1\\
0&0&1&0\\
0&1&0&0\\
-1&0&0&0
\end{pmatrix}\,,
\quad
\ga_2=
\begin{pmatrix}
0&~0~&~0~&i\;\\
0&0&i&0\\
0&-i&0&0\\
-i&0&0&0 
\end{pmatrix}\,, 
\quad 
\ga_3=
\begin{pmatrix}
0~&~0~&~1~&~0\\
0&0&0&1\\
1&0&0&0\\
0&1&0&0
\end{pmatrix}\,, 
\nln  
& \ga_0=
\begin{pmatrix}
0&~0~&~1~&0\\
0&0&0&-1\\
-1&0&0&0\\
0&1&0&0
\end{pmatrix}\,, 
\quad 
\ga_5= -i\ga_0\ga_1\ga_2\ga_3=   
\begin{pmatrix}
\;1~&~0~&0&0\\
0&1&0&0\\
0&0&-1&0\\
0&0&0&-1
\end{pmatrix}\,.
\end{align}
One can introduce $n_{\mu\nu}$ and $n_{\mu 5}$ defined as 
\begin{equation}
n_{\mu\nu} \equiv \tfrac{1}{4}[\ga_\mu,\ga_\nu]\,, \qquad 
n_{\mu5} \equiv \tfrac{1}{4}[\ga_\mu,\ga_5]\,. 
\end{equation}
Then the Lie algebra $\alg{so}(2,4)$ is spanned by the above quantities, 
\begin{align}
\alg{so}(2,4)&= {\rm span}_{\mathbb{R}}\{~\ga_\mu\,,\ga_5\,,
n_{\mu\nu}\,, n_{\mu5}~|~~\mu,\nu=0,1,2,3~\}\,,
\end{align}
and the subalgebra $\alg{so}(1,4)$ is generated by 
\begin{align}
\alg{so}(1,4)&= {\rm span}_{\mathbb{R}}\{~
n_{\mu\nu}\,, n_{\mu5}~|~~\mu,\nu=0,1,2,3~\}\,.
\end{align}

\medskip 

Now we can compute the metric of \AdS5 with Poincar\'e coordinates 
by using a coset representative, 
\begin{equation}
g=\exp\Bigl[ \,p_0\, x^0 + p_1\, x^1 + p_2\, x^2 + p_3\, x^3 \,\Bigr]\, 
\exp\left[\gamma_5\,\frac{1}{2}\log z\right]\,, 
\label{para-AdS}
\end{equation}
where the $p_{\mu} ~(\mu=0,1,2,3)$ are defined as
\begin{align}
p_{\mu} \equiv \frac{1}{2}(\ga_{\mu}-2n_{\mu5})\,.
\end{align}
The left-invariant one-form \(A = g^{-1}dg\) can now be evaluated easily. The metric is defined as 
\begin{equation}
ds^2 = g_{MN}dx^M dx^N = \Tr(A P(A))\,,
\end{equation}
where $\overline{P}$ is a coset projector from 
$\alg{so}(2,4)$ to $\alg{so}(2,4)/\alg{so}(1,4)$ 
and is defined as 
\begin{equation}
\begin{aligned}
\overline{P}(x)  &\equiv \ga_{0}\,\frac{\Tr(\ga_{0}x)}{\Tr(\gamma_0^2)}
+\sum_{i=1}^3\ga_{i}\,\frac{\Tr(\ga_{i}x)}{\Tr(\gamma_i^2)}
+\ga_{5}\,\frac{\Tr(\ga_{5}x)}{\Tr(\gamma_5^2)} 
\\
&= \frac{1}{4} \Bigl[-\ga_{0}\,\Tr(\ga_{0}x) + \sum_{i=1}^3\ga_{i}\, \Tr(\ga_{i}x)
+\ga_{5}\,\Tr(\ga_{5}x)
\Bigr] \qquad \text{for }x \in \alg{so}(2,4)\,. \label{projector}
\end{aligned}
\end{equation}
The resulting metric is given by 
\begin{equation}
ds^2 = \frac{-(dx^0)^2 + \sum_{i=1}^3(dx^i)^2 + dz^2}{z^2}\,.
\end{equation}
Note that the \AdS{} radius is set to 1\,. 
\subsection{A conformal embedding of 4D Minkowski spacetime}\label{sec:conf-emb}

Here we are interested in a coset construction of 4D Minkowski spacetime. 
However, there is the obvious problem that the standard bilinear form degenerates 
if we naively employ the usual coset $ISO(1,3)/SO(1,3)$\,. 
To avoid this, it is convenient to represent 4D Minkowski spacetime instead
as a slice of \AdS5 in Poincar\'e coordinates. 

\medskip 

A possible representation of the group element $g$ is the following:  
\begin{equation}
g=\exp\Bigl[ \,p_0\, x^0 + p_1\, x^1 + p_2\, x^2 + p_3\, x^3 \,\Bigr]\,. 
\label{para}
\end{equation}
In comparison to~(\ref{para-AdS}), the radial coordinate $z$ is not 
contained. In other words, to express 4D Minkowski spacetime,  
a section at $z=1$ of \AdS5 has been taken at this stage:  
\begin{align}
4\text{D Minkowski spacetime} = \AdS5 \big|_{z=1}\,. 
\end{align}

\medskip 

Note here that the 4D Poincar\'e algebra $\alg{iso}(1,3)$ 
and the 4D Lorentz algebra $\alg{so}(1,3)$ are generated by the following generators, 
respectively, 
\begin{equation}
\begin{aligned}
  \alg{iso}(1,3)&= {\rm span}_{\mathbb{R}}\{~n_{\mu\nu}\,,p_\mu~|~\mu,\nu=0,1,2,3~\}\,, \\
  \alg{so}(1,3)&= {\rm span}_{\mathbb{R}}\{~n_{\mu\nu}~|~\mu,\nu=0,1,2,3~\}\,. 
\end{aligned}
\end{equation}
Thus it makes sense to use the generators $p_\mu$ to parameterize 
the coset representative of $ISO(1,3)/SO(1,3)$ as~\eqref{para}\,.  
Eventually the left-invariant one-form 
$A = g^{-1}dg$ is written as a linear combination of $p_{\mu}$\,. 

\medskip 

By dropping $\ga_5$ of $\overline{P}$ in~\eqref{projector}\,,  
we introduce the projector for 4D Minkowski spacetime by  
\begin{align}
P(x)= \frac{1}{4} \Bigl[-\ga_{0}\,\Tr(\ga_{0}x) + \sum_{i=1}^3\ga_{i}\, \Tr(\ga_{i}x) \Bigr] 
\qquad \mbox{for}\quad x\in \alg{so}(2,4)\,. 
\label{Proj}
\end{align}
Then, it is straightforward to compute the metric, 
\begin{equation}
ds^2 = \Tr(AP(A)) = -(dx^0)^2 + \sum_{i=1}^3(dx^i)^2\,. \label{flat-metric}
\end{equation}
This result is the starting point of our argument in the following. 

\medskip 

The definition of the projector~\eqref{Proj} is justified as follows. 
From an algebraic point of view, we are able to 
realize 4D Minkowski spacetime as an embedded coset 
by considering the quotient of 
$\alg{so}(2,4)/\alg{so}(1,4)$ by $\ga_5$\,. That is 
\begin{align}
\frac{\alg{iso}(1,3)}{\alg{so}(1,3)}
= \frac{\alg{so}(2,4)}{\alg{so}(1,4)\oplus {\rm span_{\mathbb{R}}\{\ga_5\}}}\,,  
\label{emb}
\end{align}
where the bilinear form of $\alg{so}(2,4)$ is {\it not} degenerate. 
On the right-hand side, the appropriate coset projector turns out to be  
\begin{align}
P : ~~\alg{so}(2,4) ~\longrightarrow~ 
\frac{\alg{so}(2,4)}{\alg{so}(1,4)\oplus {\rm span_{\mathbb{R}}\{\ga_5\}}}\,.  
\end{align}
This is the reason why the 4D Minkowski projector $P$ is given by~\eqref{Proj} 
instead of the \AdS5 coset projector $\overline{P}$ in~\eqref{projector}\footnote{
Actually, also the projector  $\overline{P}$ would lead to 
the Minkowski metric~(\ref{flat-metric}) but this seems to be accidental.}. 

\medskip 

The point is that we consider a conformal embedding of 
4D Minkowski spacetime as in Eq.~\eqref{emb}
and use the coset projector $P$ in~(\ref{Proj}) 
to avoid the degeneracy of the bilinear form of 
the 4D Poincar\'e group $ISO(1,3)$\,. 
After that, the projected one-form $P(A)$ is expanded 
in terms of $\gamma_{\mu}$ ($\mu=0,1,2,3$) and the trace operation
in the action leads to non-vanishing quantities 
as we will see in the next subsection. 

\subsection{Yang--Baxter sigma model for 4D Minkowski spacetime}\label{sec:YB-4D-Min}

Yang--Baxter deformations have only been discussed for curved backgrounds so far. 
However, it is possible to generalize the formulation to Minkowski spacetime. 

\medskip 

The deformed action is given by
\begin{equation}
\label{action}
S=-\frac{1}{2}\int_{-\infty}^{\infty}\!\!d\tau\int_{0}^{2\pi}\!\!d\sigma\,
(\gamma^{\alpha\beta}-\epsilon^{\alpha\beta})\, 
\Tr\Biggl[A_{\alpha}P\circ\frac{1}{1-2\eta R_{g}\circ P}(A_{\beta})\Biggr]\,, 
\end{equation}
where $A_{\alpha} = g^{-1}\partial_{\alpha}g$ and $g$ is given in Eq.~(\ref{para})\,. 
Here $\eta$ is a constant parameter and 
the action~(\ref{action}) is reduced to the undeformed one for $\eta=0$\,. 
The base space is 2D Minkowski spacetime with the metric 
$\gamma_{\alpha\beta}={\rm diag}(-1,1)$\,.  
The anti-symmetric tensor $\epsilon^{\alpha\beta}$ 
is normalized as $\epsilon^{\tau\sigma}=1$\,. 
The operator $R_{g}$ is defined as 
\begin{equation}
R_{g} \equiv g^{-1}R(gXg^{-1})g\,,
\end{equation}
where a linear operator $R : \alg{so}(2,4)\to \alg{so}(2,4)$ 
is a solution of the \textsc{cybe}\,,
\begin{equation}
\bigl[R(M),R(N)\bigr]-R\left([R(M),N]+[M,R(N)]\right)=0\,,\qquad M, N\in \alg{so}(2,4)\,. 
\label{CYBE-2}
\end{equation}
The $R$-operator is related to the {\it skew-symmetric} classical $r$-matrix 
in the tensorial notation through
\begin{equation}
R(X)=\Tr_{2}[r(1\otimes X)]=\sum_{i}(a_{i}\Tr(b_{i}X)-b_{i}\Tr(a_{i}X))\,, 
\end{equation}
where the classical $r$-matrix is given by 
\begin{equation}
r=\sum_{i}a_{i}\wedge b_{i}\equiv \sum_{i}(a_{i}\otimes b_{i}-b_{i}\otimes a_{i})\,
\end{equation}
satisfying the \textsc{cybe}, 
\begin{equation}
\left[r_{12},r_{13}\right] +\left[r_{12},r_{23}\right]+\left[r_{13},r_{23}\right]=0\,. \label{CYBE-1}
\end{equation}
The generators $a_{i}, b_{i}$ are elements of $\alg{so}(2,4)$. 
This means that the Yang--Baxter deformations are investigated 
within $\alg{so}(2,4)$\,. 

\medskip 

It is easy to extend this formulation to the m\textsc{cybe}, which will be discussed in Section~\ref{sec:Drinfeld-Jimbo}. 

\medskip 

It is worth remarking on the order of the operations of taking a slice at $z=1$ and performing a Yang--Baxter 
deformation. Naively, it seems likely that the two operations are commutative. 
This is however not the case. The deformation may depend on the radial $z$-direction. 
For example, if the classical $r$-matrix contains $\gamma_5$\,, which deforms the $z$-direction, 
then the two operations do not commutate. On the other hand, they commute 
for all the examples listed in Sec.~\ref{sec:trivial-r}. The magnetic case for example corresponds to 
a slice of the gravity dual for non-commutative gauge theory~\cite{HI,MR} 
and in fact the classical $r$-matrix is identical to the one found in~\cite{MR-MY}. 

\subsection{Embedding into string theory}\label{sec:emd-string}

Starting from a 2D integrable model, it is natural to ask whether it can be embedded into string theory. 
This requires in general supplementing the fields we can read off from the action by other fields 
in order to solve the one-loop beta function of string theory. 
As a result of Klimcik's procedure, we always get a 4D metric 
and $B$--field. For the missing internal directions, we make the simplest ansatz 
of assuming them to be flat, as resulting from an internal $T^6$ with vanishing $B$--field. 
Adding only a dilaton will be sufficient to produce a solution for a large class of deformations, 
which all turn out to be related via T--duality to flat space. 
The minimal ansatz for the dilaton is that it preserves the same symmetries as the known fields. 
When the $r$-matrix only contains $p_\mu$ or $n_{\mu\nu}$\,, this is sufficient to find a solution. 
More general $r$-matrices which contain also $\gamma_5$ require also Ramond--Ramond fields 
in order to solve the equations. 
We have not actually been able to find the string theory embedding for all of these cases.

\section{$TsT$--duals of flat space}\label{sec:TsT-flat-space}

In this section, we have collected a class of models for which the simplest ansatz 
for a string theory embedding can be used, in which we assume the extra six dimensions 
to be flat (as resulting from a $T^6$--compactification) and we introduce only a dilaton field. 
They all turn out to be $TsT$--transformations of flat space with identifications. 
In cases in which the $r$-matrix contains a $p_\mu$, one or both T--dualities act trivially.

\medskip

We will present the examples in the following order. We start with a rather general example 
containing both $p$ and $n$ in the $r$-matrix, which corresponds to the T--dual of Melvin background. 
The $r$-matrices of the next two examples presented in Sec.~\ref{sec:T-Grant} and Sec.~\ref{sec:simple} 
have the same structure of $p\wedge n$ and correspond to the T--dual of Grant space 
and T--duals of Minkowski spacetime. The next example, the pp-wave background in Sec.~\ref{sec:pp-wave}, 
has a more complicated $r$-matrix with the structure $r\sim(p_{0}-p_{3})\wedge n$. 
Next we discuss the Hashimoto--Sethi background which has two $n$s appearing in the $r$-matrix, 
$r\sim p_{\mu}\wedge(n_{\nu\rho}+n_{\sigma\gamma})$. The next two examples have again a simpler structure of the $r$-matrix, 
with the Spradlin--Takayanagi--Volovich background, where only two $n$'s appear in Sec.~\ref{sec:STV}, 
and with models with only $p$'s in the $r$-matrix appearing in Sec.~\ref{sec:trivial-r}. 

\medskip

All the models considered in this section can be embedded in string theory 
as \(TsT\)--duals of flat space. Let \(r = a \wedge b\) and 
let \(\alpha\) and \(\beta\) be the dual coordinates to \(a\) and \(b\) 
in the sense of the Lie algebra (concretely, if \(a = p_\mu\) then \(\alpha = x^\mu\) 
and if \(a = n_{\mu\nu}\) then \(\alpha\) is the angle in the plane \(\langle x^\mu, x^\nu \rangle\)). 
Impose periodic boundary conditions on \(\alpha\) and \(\beta\) and 
consider the torus with parameter \(\tau\) generated by \(\langle \alpha, \beta \rangle\). 
The effect of the deformation is to transform \(\tau\) into \(\tau_\eta = \tau / (1 + \eta \tau)\). 
In string theory this is realized via the following sequence of transformations (\(TsT\)):
\begin{enumerate*}
\item \(T\)--dualize \(\alpha \to \tilde \alpha\); 
\item shift \(\beta \to \eta \tilde \alpha + \beta\);
\item \(T\)--dualize \(\tilde \alpha \to \alpha\).\cite{LM}.
\end{enumerate*}
In the special case of $a=p_\mu$ and $b=n_{\rho\sigma}$, 
this corresponds to the Melvin twist where $x^\mu$ is the Melvin circle.

\subsection{T--dual of Melvin Background}\label{sec:Hashimoto-Thomas}

Here we will consider an explicit example of a Yang--Baxter deformation.  
Our first example is the classical $r$-matrix
\begin{align}
r=\frac{1}{2}\,p_{3}\wedge n_{12}\,.
\end{align}
The associated geometry is a Melvin twist of 4D Minkowski spacetime. 

\medskip

Let us first compute the explicit form of $A_{\alpha}$ and a deformed current $J_{\alpha}$ 
defined as 
\begin{align}
J_{\alpha}\equiv \frac{1}{1-2\eta R_{g}\circ P}\,A_{\alpha}\,.
\end{align}
First of all, $P(A_{\alpha})$ is evaluated as  
\begin{align}
\label{pcur}
P(A_{\alpha}) = 
\frac{1}{2}\left[
\ga_{0}\, \partial_{\alpha}x^0 
+\ga_{1}\, \partial_{\alpha}x^1
+\ga_{2}\, \partial_{\alpha}x^2
+\ga_{3}\, \partial_{\alpha}x^3
\right] \,. 
\end{align}
Note here that $P(A_{\alpha})$ can also be expressed as 
\begin{equation}
\label{peqn}
\begin{aligned}
  P(A_{\alpha}) &= P\circ\Bigl(1-2\eta R_{g}\circ P\Bigr)(J_{\alpha}) \\ 
                &=P(J_{\alpha})-2\eta P\circ R_{g}(P(J_{\alpha}))\,.
\end{aligned}
\end{equation}
Then, by plugging~(\ref{pcur}) into~(\ref{peqn}) and solving the four equations, 
the explicit form of $P(J_{\alpha})$ can be determined to be
\begin{equation}
P(J_{\alpha})=\ga_{0}\, j_{\alpha}^0 + \ga_{1}\, j_{\alpha}^1 + \ga_{2}\, j_{\alpha}^2+  \ga_{3}\,j_{\alpha}^3\,,
\end{equation}
where the components of $j_{\alpha}^{\mu}$ are given by 
\begin{equation}
  \begin{aligned}
    j_{\alpha}^0 &= \frac{1}{2}\partial_\alpha x^0\,,  \\
    j_{\alpha}^1 &= \frac{\partial_\alpha
      x^1+\eta^2x^1(x^1\partial_\alpha x^1+x^2\partial_\alpha x^2)
      -\eta x^2\partial_\alpha x^3}{2(1+\eta^2[(x^1)^2+(x^2)^2])}\,, \\
    j_{\alpha}^2 &= \frac{\partial_\alpha
      x^2+\eta^2x^2(x^1\partial_\alpha x^1+x^2\partial_\alpha x^2)
      +\eta x^1\partial_\alpha x^3}{2\left(1+\eta^2\left[(x^1)^2+(x^2)^2\right]\right)}\,, \\
    j_{\alpha}^3 &= \frac{\partial_\alpha x^3+\eta
      (x^2 \partial_\alpha x^1 -x^1\partial_\alpha
      x^2)}{2(1+\eta^2[(x^1)^2+(x^2)^2])}\,.
  \end{aligned}
\end{equation}
Thus the classical action can be rewritten as
\begin{equation}
  \begin{aligned}
    S ={}& -\frac{1}{2}\int_{-\infty}^\infty\!
    d\tau\int_{0}^{2\pi}\!\!d\sigma\, \gamma^{\alpha
      \beta}\Bigl[-\partial_{\alpha}x^0\partial_{\beta}x^0+\partial_{\alpha}r\partial_{\beta}r
    +\frac{r^2\partial_{\alpha}\theta\partial_{\beta}\theta
      +\partial_{\alpha}x^3\partial_{\beta}x^3}{1+\eta^2r^2}\Bigr]  \\
    & +\int_{-\infty}^\infty\! d\tau\int_{0}^{2\pi}\!\!d\sigma\,
    \epsilon^{\alpha\beta}\frac{\eta
      r^2}{1+\eta^2r^2}\partial_{\alpha}\theta\partial_{\beta} x^3\,,
  \end{aligned}
\end{equation}
where we have performed a coordinate transformation,
\begin{align}
  x^1 &= r\,\cos\theta\,, &x^2 &= r\,\sin\theta\,.
\end{align}
From this action, one can read off the metric and $B$--field, 
\begin{equation}
  \begin{aligned}
    ds^{2} &=-(dx^0)^2+dr^{2}
    +\frac{r^2d\theta^2+(dx^3)^2}{1+\eta^2r^2}\,,
     \\
    B&= \frac{\eta r^2}{1+\eta^2r^2}\,d\theta \wedge dx^3\,.
  \end{aligned}
\end{equation}

It should be remarked that the Yang--Baxter deformations cannot reproduce the associated dilaton, 
although it may be possible to perform a supercoset construction in principle. 
However we can embed the background in string theory observing that the one-loop beta function vanishes 
by adding a dilaton~\cite{Gibbons:1987ps,Tseytlin:1994ei,HT}
\begin{equation}
\Phi = -\frac{1}{2} \log (1+\eta^2r^2)\,.
\end{equation}
The background obtained in this way is a \(TsT\) transformation of
flat space on the torus generated by \(x^3\) and \(\theta\),
\emph{i.e.} the result of the following chain of transformations:
\begin{enumerate*}
\item \(T\)--duality in \(x^3\);
\item shift \(\theta \to \eta \tilde x^3 + \theta\);
\item \(T\)--duality in \(\tilde x^3\).
\end{enumerate*}
Consistency requires \(\tilde x^3\) to be periodic with period
\(\tilde x^3  \simeq \tilde x^3 + 2 \pi / \eta\) and \(x^3\) with period
\(x^3 \simeq x^3 + \alpha' \eta/(2 \pi)\).

\subsection{Generalized Melvin Backgrounds}

\subsubsection{T--dual of Grant space}\label{sec:T-Grant}

Let us consider the classical $r$-matrix 
\begin{equation}
 r=\frac{1}{2}p_{1}\wedge n_{03}\,.
\end{equation}
The derivation is almost the same as in the previous subsection, 
hence we will not repeat the detailed explanation but simply present 
the deformed metric and $B$--field. 
The associated metric and $B$--field are given by 
\begin{equation}
  \label{Tai}
  \begin{aligned}
    ds^2 &=
    -\frac{2dx^-dx^++\eta^2(x^+dx^-+x^-dx^+)^2}{1+2\eta^2\,x^-x^+}
    +\frac{(dx^1)^2}{1+2\eta^2\,x^-x^+}+(dx^2)^2 \\
     &=    -dt^2+(dx^2)^2+\frac{1}{1+\eta^{2}\,t^2}\bigl[(dx^1)^2+t^2d\phi^2\bigr]\,,
     \\
    B &= \frac{\eta}{1+2\eta^2\,x^-x^+}(x^+dx^--x^-dx^+) \wedge dx^1 \\
    &=  \frac{\eta\, t^2}{1+\eta^2\,t^2}\,dx^1\wedge
    d\phi\,. 
  \end{aligned}
\end{equation}
Here the light-cone coordinates are given by
\begin{align}
  x^\pm \equiv \frac{x^0\pm x^3}{\sqrt{2}}\,,
\label{LC}
\end{align}
and we have introduced
new coordinates given by
\begin{align}
  \label{new}
  x^0 &=t\, \cosh \phi\,, & x^3 &= t\, \sinh \phi\,. 
\end{align}
The metric and $B$--field in~(\ref{Tai}) agree with  (2.7) and (2.8) in~\cite{Tai} for $x^2=0$\,. 

The background can be embedded in string theory by adding a dilaton
\begin{equation}
\Phi = -\frac{1}{2} \log(1+\eta^2 t^2)\,, 
\end{equation}
which solves the beta function equations.
In fact this is an exact string theory, resulting from a \(TsT\) transformation of flat space 
on the torus generated by \(x^1\) (which has periodicity \(\alpha' \eta/(2\pi)\)) and \(\phi\).

\subsubsection{Time-like Melvin circle}\label{sec:simple}

Let us consider the classical $r$-matrix
\begin{equation}
r=\frac{1}{2}\, p_{0}\wedge n_{12} \,.
\end{equation} 
This $r$-matrix is Abelian and the associated metric and $B$--field are given by 
\begin{equation}
  \begin{aligned}
    ds^2 &= -\frac{(dx^0)^2}{1-\eta^2r^2} + dr^2 +
    \frac{r^2d\theta^2}{1-\eta^2r^2}+(dx^3)^2\,, \\
    B &= \frac{\eta r^2}{1-\eta^2r^2}\, dx^0 \wedge d\theta\,.
  \end{aligned}
\end{equation}
The Ricci scalar curvature is negative, 
\begin{equation}
  R = -\frac{2\eta^2(5+2r^2\eta^2)}{(1-\eta^2 r^2)^2},
\end{equation}
and this background has a curvature singularity at $r=1/\eta$. 
This background is related to the previous one by analytic continuation 
\begin{equation}
(t, x^1, \phi)\to(ir, ix^0, i\theta).
\end{equation}
By supplementing the fields by a dilaton 
\begin{equation}
\Phi = - \frac{1}{2} \log (1-\eta^2 r^2)
\end{equation}
which solves the beta function equations, we obtain a $TsT$--dual of flat space 
on the torus $\langle x^0, \theta\rangle$. 
Note that the Melvin circle $x^0$ has periodicity $x^0 \simeq x^0 + \alpha'\eta/(2\pi)$.

\subsection{PP-wave background}\label{sec:pp-wave}

Let us consider the classical $r$-matrix 
\begin{equation}
r=\frac{1}{2\sqrt{2}}(p_{0}-p_{3})\wedge n_{12}\,. \label{1st}
\end{equation}
The associated deformed background is
\begin{equation}
  \label{DN1}
  \begin{aligned}
    ds^{2}&=-2dx^+dx^--\eta^2r^2(dx^+)^2+(dr)^2 +r^2 d\theta^2\,, \\
    B &= \eta r^2\,dx^+\wedge d\theta\,, 
  \end{aligned}
\end{equation}
where we have introduced the polar coordinate system for $x^1$ and $x^2$ given by 
\begin{align}
x^1 = r \cos\theta\,, \qquad x^2 = r \sin\theta\,.
\label{polar12} 
\end{align} 
This is a pp-wave background which can also be understood as a generalization of a \(TsT\) transformation obtained as
\begin{enumerate*}
\item a T--duality from \(\theta\) to \(\tilde \theta\), followed by 
\item the shifts \(x^0 \to \eta \tilde \theta + x^0\), \(x^3 \to - \eta \tilde \theta + x^3\) and the final
\item T--duality from \(\tilde \theta \) to \(\theta\).
\end{enumerate*}

\medskip

Note that this requires the identifications \(x^0 \simeq x^0 + \alpha'
\eta/(2 \pi)\) and \(x^3 \simeq x^3 + \alpha' \eta/(2 \pi)\).

\subsection{Hashimoto--Sethi background}\label{sec:Hashimoto-Sethi}

We next consider the classical $r$-matrix
\begin{equation}
r=\frac{1}{2\sqrt{2}}\,p_{2}\wedge(n_{01}+n_{13}) \,.
\end{equation}
The resulting metric and $B$--field are given by
\begin{equation}
  \label{HS}
  \begin{aligned}
    ds^2 &= -2dx^-dx^++\frac{1}{1+\eta^2(x^+)^2}((dx^1)^2+(dx^2)^2
    +\eta^2x^1dx^+(2x^+dx^1-x^1dx^+))\,,\\
    B &= \frac{\eta}{1+\eta^2 (x^+)^2}(x^1dx^+-x^+dx^1)\wedge
    dx^2\,. 
  \end{aligned}
\end{equation}
\medskip 
Note that this background depends on the light-cone time $x^+$ explicitly. 
The associated dilaton to complete the string embedding is taken to be
\begin{equation}
\Phi = -\frac{1}{2} \log (1 + \eta^2 (x^+)^2)\,.
\end{equation}

\medskip

The metric and $B$--field~(\ref{HS}) agree with those of the Hashimoto--Sethi background. 
To show this agreement, one has to introduce new coordinates, 
\begin{align}
x^+ &=\frac{1}{\eta}y^+\,, &
x^1 &=y^+{\tilde y}\,, &
x^- &=\eta y^-+\frac{\eta}{2}y^+{\tilde y}^2\,,  &
x^2 &=-{\tilde z}\,.
\end{align}
This reproduces the expression in Eq.~(25) of~\cite{HS} 
where the background is shown to be the result of a \(TsT\) transformation.

\subsection{Spradlin--Takayanagi--Volovich background}\label{sec:STV}

Let us here consider the classical $r$-matrix
\begin{equation}
r=\frac{1}{2}n_{12}\wedge n_{03} \,.
\end{equation}
Then the associated metric and $B$--field are given by
\begin{equation}
 \label{STV-0}
  \begin{aligned}
    ds^2 &= -2dx^+dx^-+dr^2+\frac{1}{1+2\eta^2r^2x^-x^+}\bigl(r^2d\theta^2-r^2\eta^2
    (x^+dx^--x^-dx^+)^2\bigr)\,, \\
    B &= \frac{\eta r^2}{1+2\eta^2r^2x^-x^+}(x^-dx^+-x^+dx^-)\wedge
    d\theta\,.
  \end{aligned}
\end{equation}
The light-cone coordinates are given in~(\ref{LC})\,. 
This background (\ref{STV-0}) is really time-dependent and 
has a curvature singularity. 

\medskip 

By using the coordinates in~(\ref{new}), one can rewrite the expressions in (\ref{STV-0}) as 
\begin{equation}
\label{STV}
  \begin{aligned}
    ds^2 &= -dt^2+dr^2+\frac{r^2d\theta^2+t^2d\phi^2}{1+\eta^2r^2t^2}\,, \\
    B &= \frac{\eta r^2t^2}{1+\eta^2r^2t^2}d\phi \wedge d\theta\,. 
  \end{aligned}
\end{equation}
Note that the coordinates in (\ref{new}) do not cover the whole $x^0$-$x^3$ plane  
and the background (\ref{STV}) contains no singularity. 
Then the metric and $B$--field in~(\ref{STV}) agree with those of (6.1) in~\cite{STV}.  
This is a time-dependent background realized 
by a \(TsT\) transformation of Minkowski spacetime on the torus generated by \(\phi\) and \(\theta\).
The associated dilaton is 
\begin{equation}
\Phi = -\frac{1}{2} \log(1 + \eta^2 r^2 t^2)\,.
\end{equation}

\subsection{Locally flat spaces}\label{sec:trivial-r}

In this section, we consider classical $r$-matrices of the type 
\begin{equation}
r=\frac{1}{2}\,p_{\mu}\wedge p_{\nu} \qquad (\mu,\nu=0,1,2,3)\,.
\end{equation}
In these cases the associated geometries are locally flat and the \(B\)--field is closed. 
The backgrounds are nevertheless non-trivial and are associated to non-commutative field theories~\cite{SW}.
The following is a list of possible $r$-matrices. 

\begin{description}

\item[(1) Magnetic flux] \qquad $r=\frac{1}{2}\,p_{i}\wedge p_{j}~~(i\neq j,~i,j=1,2,3)$ \qquad  (Melvin Shift Twist)

With this $r$-matrix, the following metric and $B$--field are obtained: 
\begin{equation}
  \begin{aligned}
    ds^2 &= -(dx^0)^2+\frac{(dx^i)^2+(dx^j)^2}{1+\eta^2}
    +\sum_{k=1}^3 \ep^{ijk}(dx^k)^{2} \, ,\\
    B &= \frac{\eta}{1+\eta^2}dx^i\wedge dx^j\,.
  \end{aligned}
\end{equation}
This background is obtained via a \(TsT\) transformation on the torus \(\langle  x^i, x^j \rangle\).

\medskip 

A multi-parameter generalization of this type is obtained using the $r$-matrix 
\begin{align}
r=\frac{1}{2}(a_3\, p_1\wedge p_2 +a_1\, p_2\wedge p_3 +a_2\, p_3\wedge p_1 )\,, 
\end{align}
where $a_1,a_2,a_3 \in \mathbb{R}$ are deformation parameters. 
The resulting metric and $B$--field are given by 
\begin{equation}
\label{multi-mag}
  \begin{aligned}
    ds^2&= -(dx^0)^2+
    \frac{(dx^1)^2+(dx^2)^2+(dx^3)^2+\eta^2(a_1dx^1+a_2dx^2+a_3dx^3)^2}
    {1+\eta^2 (a_1^2+a_2^2+a_3^2)}\,, \\ 
    B&=\frac{\eta}{1+\eta^2 (a_1^2+a_2^2+a_3^2)} \left(a_3\,
      dx^1\wedge dx^2+a_1\, dx^2\wedge dx^3+a_2\,dx^3\wedge dx^1
    \right) \,.
  \end{aligned}
\end{equation}

\item[(2) Electric flux] \qquad $r=\frac{1}{2}\,p_{0}\wedge p_{i}~~(i=1,2,3)$ \qquad  (Melvin Shift Twist)

This $r$-matrix leads to the metric and $B$--field
\begin{equation}
  \begin{aligned}
    ds^{2} &= \frac{-(dx^0)^2+(dx^i)^2}{1-\eta^2}
    +\sum_{ j\neq i}(dx^j)^2 \,, \\
    B &=  -\frac{\eta}{1-\eta^2}\,dx^0\wedge dx^i\,.
  \end{aligned}
\end{equation}
This corresponds to turning on an electric flux and the range of $\eta$ is restricted to $-1<\eta < 1$\,. 
Once more we can obtain this background via a \(TsT\) transformation on the torus \(\langle x^0, x^i \rangle\).
The background is related to the non-commutative open string theory (\textsc{ncos})~\cite{NCOS}.

\medskip

A multi-parameter generalization of this type is obtained by the $r$-matrix 
\begin{align}
r=\frac{1}{2}\, p_0 \wedge (a_1\, p_1 +a_2\, p_2 +a_3\, p_3 )\,, 
\end{align}
where $a_1,a_2,a_3 \in \mathbb{R}$ are deformation parameters. 
The resulting metric and $B$--field are given by 
\begin{align}
ds^2&= 
\frac{
-(dx^0)^2+\eta^2(a_1dx^1+a_2dx^2+a_3dx^3)^2 }
{1-\eta^2(a_1^2+a_2^2+a_3^2 )}
+(dx^1)^2+(dx^2)^2+(dx^3)^2   
\,, \el \\
B&=\frac{\eta}{1-\eta^2 (a_1^2+a_2^2+a_3^2)}
\left(a_1\, dx^1+a_2\, dx^2+a_3\,dx^3\right)\wedge dx^0 \,. 
\end{align}

\item[(3) Light-like flux]  \qquad $r=\frac{1}{2\sqrt{2}}\,(p_{0}-p_{3})\wedge p_{2}$ \qquad (Null Melvin Shift Twist) 

In this case, we obtain the following metric and $B$--field 
\begin{equation}
  \begin{aligned}
    ds^{2} &=-2dx^+dx^--\eta^2(dx^+)^2+(dx^1)^2+(dx^2)^2\,,\\
    B &= - \eta\, dx^+\wedge dx^2\,.
  \end{aligned}
\end{equation}
This corresponds to turning on a light-like flux resulting from a \(T\)--duality in \(x^2\), 
opposite shifts in \(x^0\) and \(x^3\) and a final \(T\)--duality in \(\tilde x^2\).
The background is related to the light-like non-commutativity argued in~\cite{AGM}. 
\end{description}

\section{Non-Twist cases }\label{sec:dS-AdS}

So far, we have discussed the cases associated with $TsT$ transformations. 
In this section, we shall consider the other cases, for instance, 
including a single T--duality, or a chain of $TsT$-transformations 
followed by an S--duality. We will refer to the cases as non-twist cases. 

\medskip 

As the first example, we revisit the pp-wave case 
and argue its relation to S--duality. 
Then we will reproduce T--duals of \dS4 and \AdS4 respectively, 
as Yang--Baxter deformations. 
These examples may be regarded as non-trivial examples of our procedure. 
Finally we provide more complicated examples including two-parameter deformations. 

\medskip 

Let us remark that the presence of $\gamma_5$ in the all of the $r$-matrices appearing in this section 
plays a crucial role. It can intuitively be interpreted as a deformation of the radial direction 
of \AdS5 and indicates that the location of the slice is moved. 
This is also the reason why the integrability of the $TsT$--cases of the previous section is obvious, 
while in general it is not automatic in the non-twist cases, except for special cases such as the T--duals of 
\dS4 and \AdS4.

\subsection{PP-wave revisited -- S-duality}

In the previous section, we have considered the pp-wave background (\ref{DN1}). 
We will show that the same metric can be reproduced by considering 
the $r$-matrix
\begin{equation}
\label{rh3p+}
r=\frac{1}{4\sqrt{2}}(\ga_5-2n_{03})\wedge (p_{0}-p_{3})\,.
\end{equation}
The resulting metric and $B$--field are given by 
\begin{equation}
  \label{DN2}
  \begin{aligned}
    ds^2 &= -2dx^+dx^--\eta^2r^2(dx^+)^2+(dr)^2 +r^2 d\theta^2\,, \\
    B &= \eta r \,dx^+ \wedge dr \,. 
  \end{aligned}
\end{equation}
As noted above, the metric is the same as in~(\ref{DN1}), 
but the $B$--field is different. In particular, the $B$--field carries an index for the radial direction $r$\,, 
which is not a $U(1)$-direction. Hence it does not seem likely that the background~(\ref{DN2}) 
can be obtained via a simple twisting procedure, as opposed to~(\ref{DN1}). 
This difference comes from the fact that the second $r$-matrix~(\ref{rh3p+}) is 
non-Abelian, while the first one~(\ref{1st}) is Abelian. 
This non-Abelian $r$-matrix appears to be related to a chain of dualities. 
In fact, the $r$-matrix~(\ref{rh3p+}) is identical 
to the one in (2.17) of~\cite{SUGRA-KMY} that has been employed 
to study a deformation of \(\AdS5 \times S^5\).
It has been shown in~\cite{MY-duality} that this $r$-matrix 
corresponds to a duality chain of three TsT transformations and 
an S-duality\footnote{In this spirit, we can understand the background in Eq.~\eqref{DN2} 
as the S--dual of the one in Eq.\eqref{DN1}.}. 
Thus non-Abelian $r$-matrices appear to be related to intricate duality chains. 

\medskip 

As another possibility, non-Abelian $r$-matrices may indicate deformations 
beyond the standard twists or dualities. 
To elaborate this statement, it would be interesting 
to study more non-trivial examples of classical $r$-matrices, 
as we will show in the next subsection.

\subsection{T--dual of \dS4}

Now we consider the classical $r$-matrix
\begin{equation}
r=\frac{1}{4}\ga_5 \wedge p_0\,.
\end{equation}
Note that this $r$-matrix contains $\gamma_5$ and 
leads to the metric and $B$--field given by 
\begin{equation}
  \label{dS1}
  \begin{aligned}
    ds^2 
    ={}& \frac{-(dx^0)^2+dr^2}{1-\eta^2r^2}+r^2 \sin^2\theta d\phi^2+r^2d\theta^2\,,  \\
    B 
    ={}& \frac{\eta r}{1-\eta^2r^2}dx^0\wedge dr\,. 
  \end{aligned}
\end{equation}
where we have introduced new coordinates $r$\,, $\theta$ and $\phi$ through 
\begin{align}
  x^1 &= r \cos\phi\,\sin\theta\,, & x^2 &=r \sin\phi\sin\theta\,, & x^3 &= r \cos \theta\,.
\end{align}
Note here that the above $B$--field can be rewritten in the form of a total derivative. 

\medskip

The deformed background~(\ref{dS1}) is already simple.  
However, in order to understand the background well, 
we can perform a T--duality along the $x^0$-direction\footnote{This is a time-like T--duality. 
For the detail, see the original argument~\cite{Hull}.}.
The resulting background is given by 
\begin{equation}
ds^2=(dr+\eta r\,dx^0)^2-(dx^0)^2+r^2(d\theta^2+{\rm sin}^2\theta\, d\phi^2)\,.
\end{equation} 
Note that the $B$--field has disappeared now. 
Then, by performing a coordinate transformation from $x^0$ to $t$\,, 
\begin{equation}
x^0=t-\frac{1}{2\eta}{\rm log}(\eta^2r^2-1)\,, 
\end{equation}
one can reproduce the well-known metric of \dS4 in static coordinates, 
\begin{equation}
ds^2=-(1-\eta^2r^2)dt^2+\frac{dr^2}{1-\eta^2r^2}+r^2(d\theta^2+{\rm sin}^2\theta d\phi^2)
\end{equation}
with a cosmological horizon at $r=1/\eta$\,. 
Thus we have shown that the background~(\ref{dS1}) is nothing but a T--dual of \dS4. 
The \dS4 geometry cannot be realized as a twist of 4D Minkowski spacetime, 
hence it is quite remarkable that the T--dual of \dS4 has been obtained as a Yang--Baxter 
deformation. 

\subsection{T--dual of \AdS4}

As another example, let us consider the classical $r$-matrix
\begin{equation}
r=\frac{1}{4}\ga_5\wedge p_{1}\,.
\end{equation}
This $r$-matrix also contains $\gamma_5$\,. 
The associated metric and $B$--field are given by
\begin{equation}
  \begin{aligned}
    ds^2
    ={}&\frac{dt^2+(dx^1)^2}{1+\eta^2t^2}+t^2 \cosh^2\phi d\theta^2-t^2d\phi^2\,, \\
    B 
    ={}&\frac{\eta t}{1+\eta^2t^2}dt\wedge dx^1\,,
  \end{aligned}
\end{equation}
where we have introduced new coordinates $t$\,, $\theta$ and $\phi$ through 
\begin{align}
  x^0&=t\, \sinh\phi\,, &
                          x^2&=t\, \cos\theta\, \cosh\phi\,, &
                                                               x^3&=t\, \sin\theta\, \cosh\phi \,.
\end{align}
Note here that the $B$--field can be recast as a total derivative. 

\medskip

As in the previous case, it is nice to perform a T--duality along the $x^1$-direction. 
Then the resulting background is given by 
\begin{equation}
ds^2=(dt-\eta t\,dx^1)^2+(dx^1)^2+t^2(-d\phi^2 + \cosh^2\phi\, d\theta^2)\,.
\end{equation} 
Now the $B$--field has disappeared. Let us perform a coordinate transformation,  
\begin{equation}
x^1 = y + \frac{1}{2\eta}\log(\eta^2t^2+1)\,.
\end{equation}
Then the resulting metric is given by 
\begin{equation}
ds^2 = (1+\eta^2 t^2) dy^2 + \frac{dt^2}{1+\eta^2 t^2} + t^2 (-d\phi^2+\cosh^2\phi\,d\theta^2)\,.
\end{equation}
By replacing the coordinates (with a double Wick rotation) by 
\begin{equation}
y \to it\,, \quad t \to r\,, \quad \phi \to i\theta\,, \quad 
\theta \to \phi\,,
\end{equation}
one can obtain the standard metric of \AdS4 with the global coordinates
\begin{equation}
ds^2 = -(1+\eta^2r^2)dt^2 + \frac{dr^2}{1+\eta^2r^2}+r^2 (d\theta^2 + \cos^2\theta\,d\phi^2)\,.
\end{equation}
Note that $\eta^2$ measures the curvature.

\subsection{More complicated backgrounds}\label{sec:more-compl}

Here, we give a list of classical $r$-matrices for which the corresponding backgrounds 
have not been identified yet. 
All of the classical $r$-matrices we consider here satisfy the \textsc{cybe} given in~(\ref{CYBE-1})\,. 
Nevertheless, this does not guarantee integrability without explicitly constructing the Lax pair. 
In the previous cases, the string embedding assured integrability via T--duality to integrable backgrounds. 
We conjecture that also the following examples are new integrable models, but at this point have no explicit proof. 

\medskip 

In the following, we assume that the light-cone coordinates are defined in (\ref{LC}) and 
$x^1$\,, $x^2$ are rewritten with the polar coordinates given in~\eqref{polar12}.

\subsubsection{One-parameter deformations}

Let us consider more complicated one-parameter deformations. 

\medskip 

The first example is 
\begin{equation}
r=\frac{1}{4\sqrt{2}}\, (\ga_5-2n_{12})\wedge (p_{0}-p_{3})\,,
\end{equation}
and the resulting metric and $B$--field are given by 
\begin{equation}
  \begin{aligned}
    ds^2 &= \frac{-2dx^-dx^+ -2\eta^2 rdx^+ \bigl[ r dx^+ - x^+ (dr +
      rd\theta)\bigr]}{1-\eta^2(x^+)^2} +dr^2 + r^2 d\theta^2\,,
    \\
    B &= \frac{\eta}{1-\eta^2(x^+)^2}\, dx^+\wedge (rdr - x^+ dx^- +
    r^2 d\theta)\,.
  \end{aligned}
\end{equation}
This $r$-matrix is non-Abelian. The Ricci scalar and Riemann square vanish. 
This background has a coordinate singularity at $x^+ = \pm 1/\eta$\,.

\medskip 

The second example is 
\begin{equation}
r=\frac{1}{4}\, (n_{05}-n_{35})\wedge (p_{0}-p_{3})\,,
\end{equation}
and the metric and $B$--field are given by 
\begin{equation}
  \begin{aligned}
    ds^2 &= \frac{ -2dx^+dx^- + dr^2
      -\eta^2(x^+)^2(x^+dr-rdx^+)^2}{1-\eta^2(x^+)^4}
    + r^2 d\theta^2 \\ 
    \\
    B &= \frac{\eta x^+}{1-\eta^2(x^+)^4}\,dx^+
    \wedge(rdr-x^+dx^-)\,.
  \end{aligned}
\end{equation}
This is an Abelian $r$-matrix in which $\gamma_5$ is not contained but $n_{05}$ and $n_{35}$ carry 
the index 5. 
Note that the Ricci scalar and Riemann square vanish and 
$x^+=\pm |\eta|^{-1/2}$ are just coordinate singularities.

\subsubsection{Two-parameter deformations }\label{sec:two-param} 

It may be interesting to consider two-parameter deformations. 

\medskip

The first example is given by 
\begin{equation}
  \begin{aligned}
    r &= \frac{s_1}{2}\,E_{24}\wedge (E_{22}-E_{44}) + \frac{s_2}{2}\,E_{13} \wedge (E_{11}-E_{33}) \\
    &= \frac{s_1}{8}\,(p_0 - p_3) \wedge \left(2n_{03} - \gamma_5\right) + \frac{s_2}{8}\,(p_0 + p_3) \wedge
    \left(2n_{03} + \ga_5\right) \,,
  \end{aligned}
\end{equation}
where $s_1$ and $s_2$ are arbitrary constant parameters. This $r$-matrix has been 
used to study a two-parameter deformation of \AdS5 in~\cite{SUGRA-KMY}. 
The associated metric and $B$--field are given by 
\begin{equation}
  \begin{aligned}
    ds^2 &= \frac{-2dx^+dx^- + dr^2 }{1+s_1s_2\eta^2\,r^2}
    + r^2 d\theta^2 - \frac{\eta^2\,r^2(s_1 dx^+ + s_2 dx^-)^2}{2(1+s_1 s_2\eta^2\,r^2)}\,, \\
    B &= \frac{\eta\, r}{\sqrt{2}\,(1+s_1 s_2 \eta^2\,r^2)}\,(s_1dx^+
    - s_2 dx^-)\wedge dr\,.
  \end{aligned}
\end{equation}
When $s_1 = -s_2 = s$\,, the metric simplifies to
\begin{equation}
  \label{dS}
  \begin{aligned}
    ds^2 &= \frac{-(dx^0)^2 + dr^2 }{1-s^2\eta^2\, r^2}
    + r^2d\theta^2 + (dx^3)^2\,, \\
    B &= \frac{s\eta\,r}{1 - s^2\eta^2\,r^2}\,  dx^0 \wedge dr\,.
  \end{aligned}
\end{equation}
A T--duality along the $x^0$-direction 
leads to the geometry of $\dS3\times \mathbb{R}$\,.

\medskip

The second example is given by 
\begin{equation}
  \begin{aligned}
    r &= \frac{1}{\sqrt{2}} E_{24}\wedge \bigl[(a+ib)
    (E_{22}-\tfrac{1}{4}E)-(a-ib) (E_{44}-\tfrac{1}{4}E)\bigr]
    \\
    &= \frac{1}{4\sqrt{2}}(p_0-p_3)\wedge \bigl[a(2n_{03}-\gamma_5)+2b\,n_{12} \bigr]\,,
  \end{aligned}
\end{equation}
where $a,b\in \mathbb{R}$ are deformation parameters and 
$E \equiv \sum_{i=1}^4E_{ii}$\,. 
The resulting metric and $B$--field are given by 
\begin{equation}
  \begin{aligned}
    ds^2&= -2dx^+dx^- + dr^2  + r^2 d\theta^2  -\eta^2 (a^2+b^2) r^2 (dx^+)^2\,, \\
    B&= \eta\, r\, dx^+ \wedge (a dr+b r d\theta )\,.
  \end{aligned}
\end{equation}
Thus this background is regarded as an interpolation between the plane waves in~(\ref{DN1}) 
and~(\ref{DN2})\footnote{It would be interesting to see if this background can be understood 
as the result of an \(SL_2(\mathbb{R})\) transformation acting on the pp wave of Eq.~\eqref{DN1}.}.  

\medskip 

The third example is given by 
\begin{equation}
  \begin{aligned}
    r ={}& \frac{i}{\sqrt{2}} \Bigl[E_{24}\wedge
    \bigl(a_1(E_{22}-\tfrac{1}{4}E)-a_2 (E_{44}-\tfrac{1}{4}E)\bigr)
    -(a_1+a_2) E_{23}\wedge E_{34} \Bigr] \\
    ={}& -\frac{i}{4\sqrt{2}}\biggl[(p_0-p_3)\wedge
    \Bigl(a_1\Bigl(in_{12}-n_{03} +\frac{\gamma_5}{2} \Bigr)
    -a_2\Bigl(in_{12}+n_{03}-\frac{\gamma_{5}}{2} \Bigr)\Bigr)\\
    &+(a_1+a_2)(p_1-ip_2)\wedge
    \bigl(n_{01}+n_{13}+i(n_{02}+n_{23})\bigr)\biggr]\,,
  \end{aligned}
\end{equation}
where $a_1$\,, $a_2 \in \mathbb{R}$ are deformation parameters. 
The resulting metric and $B$--field are given by 
\begin{equation}
\begin{aligned}
  ds^2 ={}& -2dx^+dx^- + dr^2  + r^2 d\theta^2 \\ 
  &-\eta ^2\frac{ \left[(a_1+a_2)  x^+dr -a_2 r dx^+\right]^2
      +(a_1+a_2)^2 r^2(x^+)^2 d\theta^2 }{1+\eta^2 (a_1+a_2)^2 (x^+)^2}\,,  \\
  B ={}& \frac{\eta\, r}{1+\eta^2 (a_1+a_2)^2 (x^+)^2} \left((a_1+a_2) x^+ dr - a_2 r dx^+
\right)\wedge d\theta\,.
\end{aligned}
\end{equation}
When $a_1=-a_2$\,, the background in~(\ref{DN1}) is reproduced. Thus this background 
can be regarded as a deformation of the pp-wave background.

\section{A deformation of Drinfeld--Jimbo type}\label{sec:Drinfeld-Jimbo}

So far, we have considered classical $r$-matrices satisfying 
the classical Yang--Baxter equation~(\ref{CYBE-1}) (or~(\ref{CYBE-2})). 
Here, as an exceptional case, let us consider a classical $r$-matrix of Drinfeld--Jimbo (DJ) type,
\begin{equation}
r_{\rm DJ} = -i\sum_{1\leq i<j \leq 4}\,E_{ij} \wedge E_{ji}\,,\qquad (E_{ij})_{kl} 
\equiv \delta_{ik}\delta_{jl}\,, \label{DJ}
\end{equation}
which satisfies the modified Yang--Baxter equation, 
\begin{equation}
\bigl[R(M),R(N)\bigr]-R\left([R(M),N]+[M,R(N)]\right)= [M,N]\,, \label{mCYBE}
\end{equation} 
where $M, N\in \alg{so}(2,4)$.
In comparison to the \textsc{cybe} in~(\ref{CYBE-2})\,, the right-hand
side of~(\ref{mCYBE}) is modified\footnote{The $r$-matrices satisfying 
the m\textsc{cybe} in~(\ref{mCYBE}) are said to be \emph{non-split type}, 
as opposed to the \emph{split type} obtained when the right-hand side of~(\ref{mCYBE}) 
is negative. Our choice of sign ensures the reality of the $B$--field as in~\cite{DMV,ABF}.}. 
The $r$-matrix~(\ref{DJ}) was utilized to study an integrable deformation of 
the  $\AdS5 \times S^5$ superstring~\cite{DMV2}. The deformed metric and $B$--field 
were explicitly computed in~\cite{ABF}. 

\medskip 

The computation scheme is essentially identical because we do not mention the 
the associated Lax pair and the kappa transformation related to the Green--Schwarz string action. 

\medskip 

The resulting metric and $B$--field are given by 
\begin{equation}
  \label{eta}
  \begin{aligned}
    ds^2 &= -r^2\sin^2\theta\,dt^2 + dr^2 +
    \frac{r^2}{1+\eta^2\,r^4\sin^2\theta}
    \left(d\theta^2 + \cos^2\theta\,d\phi^2\right)\,, \\
    B &=
    -\frac{\eta\,r^4\sin\theta\cos\theta}{1+\eta^2\,r^4\sin^2\theta}\,d\theta\wedge
    d\phi\,.
  \end{aligned}
\end{equation}
Here we have performed a coordinate transformation,
\begin{align}
 x^0 &= r\sin\theta\,\sinh t\,, \quad\quad  
x^1 = r\cos\theta\,\cos\phi\,, \nonumber \\ 
 x^2 &= r\cos\theta\,\sin\phi\,, \quad\quad  
x^3 = r\,\sin\theta\,\cosh t\,,
\end{align}
and rescaled $\eta \to \eta/2$\,. 
It is worth noting that the metric in~(\ref{eta}) would be regular 
as opposed to the case of the deformation of \AdS5. 
The scalar curvature has no singularity. 
We are now working on a slice of \AdS5 at $z=1$ 
which is why
the singular surface of the metric identified in~\cite{ABF} has been avoided.

\section{Conclusion and Discussion}\label{sec:conclusions}

In this paper, we have discussed Yang--Baxter deformations of 
4D Minkowski spacetime using a conformal embedding of the spacetime 
into \AdS5. Via this procedure we have succeeded in reproducing 
the metric and $B$--field of  well-known backgrounds such as 
T--duals of Melvin backgrounds, Hashimoto--Sethi backgrounds, time-dependent backgrounds of 
Spradlin--Takayanagi--Volovich, the T--dual of Grant space, 
pp-wave backgrounds, and T--duals of \dS4 and \AdS4\,. For a large class of them, 
an embedding into string theory exists as a $TsT$--transformation of flat space with identifications, 
where the extra six dimensions are flat and only a dilaton is introduced. Finally we have studied 
a deformation by using a classical $r$-matrix 
of Drinfeld--Jimbo type and have derived the associated metric and $B$--field explicitly. 
The backgrounds that we have identified are summarized in Table~\ref{table}. 

\begin{table}
\begin{small}
\begin{center}
\begin{tabular}{ccc} 
\hline
\toprule
$r$-matrix & Type of Twist & Background  \\ \midrule
$ p_{i}\wedge p_{j}~(i,j=1,2,3)$ &Melvin Shift Twist  & Seiberg-Witten   \\ 
$ p_{0}\wedge p_{i}~$ &Melvin Shift Twist  & NCOS   \\ 
$(p_{0}+p_{i})\wedge p_{j}~(i\neq j)$ & ~Null Melvin Shift Twist~ & light-like NC  \\ 
$\frac{1}{2}p_{3}\wedge n_{12}$ &Melvin Twist  &T--dual Melvin  \\ 
~~$\frac{1}{2\sqrt{2}}p_{2}\wedge (n_{01}+n_{13})$~~ &Melvin Null Twist  &Hashimoto--Sethi \\ 
$\frac{1}{2}n_{12}\wedge n_{03}$ &R Melvin R Twist  & ~Spradlin--Takayanagi--Volovich~ \\ 
$\frac{1}{2}p_{1}\wedge n_{03}$ &Melvin Boost Twist  & T--dual of Grant space \\ 
$\frac{1}{2\sqrt{2}}(p_{0}-p_{3})\wedge n_{12}$ &Null Melvin Twist  & pp-wave   \\ \midrule
$\frac{1}{4\sqrt{2}}(\ga_5-2n_{03})\wedge (p_{0}-p_{3})$ & Non-Twist  & pp-wave  \\ 
$\frac{1}{4}\ga_5\wedge p_{0}$& Non-Twist &T--dual of \dS4 \\ 
$\frac{1}{4}\ga_5\wedge p_1$& Non-Twist &T--dual of \AdS4 \\ \midrule
DJ-type (m\textsc{CYBE}) & Non-Twist & $q$-deformation (?) \\ \bottomrule
\end{tabular}
\end{center}
\end{small}
\caption{\footnotesize 
A catalog of classical $r$-matrices and the associated backgrounds. 
\label{table}}
\end{table} 

\medskip 

There are many open questions. The undeformed case is trivially integrable 
and hence the integrability should be preserved under Yang--Baxter deformations. 
It would be interesting to consider what happens to the Lax pairs and to study the associated algebras. 

\medskip 

It would be most important to generalize our argument to 10D Minkowski spacetime 
in order to extend our argument to a consistent string theory. 
For this purpose, we have to consider the 10D conformal group $SO(2,10)$ 
and realize 10D Minkowski spacetime as a slice of 11D \AdS{} space. 
We expect that in this case 10D supersymmetric configurations like the fluxtrap backgrounds~\cite{HOR} 
could be reproduced as Yang--Baxter deformations. 

\medskip 

As opposed to deformations of \(\AdS5 \times S^5\), our computations are very simple 
and hence it is easier to generalize the Yang--Baxter deformations. 
A possible direction is a supercoset construction including spacetime fermions. 
Furthermore, it would be very interesting to investigate some quantum aspects 
such as the relation between string spectra and Yang--Baxter deformations.

\subsection*{Acknowledgments}

We are very grateful to Io Kawaguchi for collaboration at an early stage of this work. 
The work of K.Y.\ is supported by Supporting Program for Interaction-based Initiative Team Studies 
(\textsc{spirits}) 
from Kyoto University and by the \textsc{jsps} Grant-in-Aid for Scientific Research (C) No.\,15K05051. 
This work is also supported in part by the \textsc{jsps} Japan-Russia Research Cooperative Program and 
the \textsc{jsps} Japan-Hungary Research Cooperative Program.  

\medskip 

D.O. and S.R. would like to thank the Department of Physics of Kyoto University for hospitality. 
The work of S.R. is supported by the Swiss National Science Foundation (\textsc{snf}) under grant number PP00P2\_157571/1.

\appendix






\begin{small}

\end{small}

\begin{thebibliography}{99}

\bibitem{Luscher}
M.~L\"uscher, ``Quantum Nonlocal Charges And Absence Of Particle Production In The
Two-Dimensional Nonlinear Sigma Model,'' Nucl.\ Phys.\ B {\bf 135} (1978) 1, \\ 
M.~L\"uscher and K.~Pohlmeyer, ``Scattering Of Massless Lumps And Nonlocal Charges In
The Two-Dimensional Classical Nonlinear Sigma Model,'' Nucl.\ Phys.\ B {\bf 137} (1978) 46.

\bibitem{AAR}
  E.~Abdalla, M.~C.~B.~Abdalla and K.~Rothe, 
{\it Non-perturbative methods in two-dimensional quantum field theory}, 
Singapore: World Scientific, 1991. 

\bibitem{M}  
  J.~M.~Maldacena,
  ``The large N limit of superconformal field theories and supergravity,''
  Adv.\ Theor.\ Math.\ Phys.\  {\bf 2} (1998) 231
  [Int.\ J.\ Theor.\ Phys.\  {\bf 38} (1999) 1113]. 
  [arXiv:hep-th/9711200].

\bibitem{MT}
  R.~R.~Metsaev and A.~A.~Tseytlin,
  ``Type IIB superstring action in AdS$_5\times$S$^5$ background,''  
  Nucl.\ Phys.\ B {\bf 533} (1998) 109  [hep-th/9805028].  

\bibitem{BPR}
  I.~Bena, J.~Polchinski and R.~Roiban,
  ``Hidden symmetries of the AdS$_5\times$S$^5$ superstring,''
  Phys.\ Rev.\ D {\bf 69} (2004) 046002
  [hep-th/0305116].

\bibitem{review}
  N.~Beisert {\it et al.},
  ``Review of AdS/CFT Integrability: An Overview,'' 
  Lett.\ Math.\ Phys.\ {\bf 99} (2012) 3 [arXiv:1012.3982 [hep-th]]. 

\bibitem{Cherednik}
  I.~V.~Cherednik, 
  ``Relativistically Invariant Quasiclassical Limits Of Integrable
  Two-Dimensional Quantum Models,''
  Theor.\ Math.\ Phys.\  {\bf 47} (1981) 422
  [Teor.\ Mat.\ Fiz.\  {\bf 47} (1981) 225].

\bibitem{FR}
  L.~D.~Faddeev and N.~Y.~Reshetikhin,
  ``Integrability of the principal chiral field model in (1+1)-dimension,''
  Annals Phys.\  {\bf 167} (1986) 227.      

\bibitem{BFP}
  J.~Balog, P.~Forgacs and L.~Palla,
  ``A two-dimensional integrable axionic sigma model and T duality,''  
  Phys.\ Lett.\ B {\bf 484} (2000) 367  
  [hep-th/0004180].  







\bibitem{Klimcik}
 C.~Klimcik,
  ``Yang-Baxter sigma models and dS/AdS T duality,''  
JHEP {\bf 0212} (2002) 051  [hep-th/0210095]; 
  ``On integrability of the Yang-Baxter sigma-model,''  
J.\ Math.\ Phys.\  {\bf 50} (2009) 043508  [arXiv:0802.3518 [hep-th]]; 
 ``Integrability of the bi-Yang-Baxter sigma model,''  Lett.\ Math.\ Phys.\  {\bf 104} (2014) 1095
  [arXiv:1402.2105 [math-ph]].  


\bibitem{KY}
  I.~Kawaguchi and K.~Yoshida,
  ``Hidden Yangian symmetry in sigma model on squashed sphere,''
  JHEP {\bf 1011} (2010) 032. 
  [arXiv:1008.0776 [hep-th]]. 

\bibitem{KYhybrid}
  I.~Kawaguchi and K.~Yoshida,
  ``Hybrid classical integrability in squashed sigma models,''
  Phys.\ Lett.\ B\ {\bf 705} (2011) 251
  [arXiv:1107.3662 [hep-th]]; 
   ``Hybrid classical integrable structure of squashed sigma models: A short summary,''  
  J.\ Phys.\ Conf.\ Ser.\  {\bf 343} (2012) 012055 
  [arXiv:1110.6748 [hep-th]].    
  
\bibitem{KMY-QAA}
  I.~Kawaguchi, T.~Matsumoto and K.~Yoshida,
  ``The classical origin of quantum affine algebra in squashed sigma models,''  
  JHEP {\bf 1204} (2012) 115  [arXiv:1201.3058 [hep-th]]; 
  ``On the classical equivalence of monodromy matrices in squashed sigma model,''  
  JHEP {\bf 1206} (2012) 082  [arXiv:1203.3400 [hep-th]].


\bibitem{DMV}
  F.~Delduc, M.~Magro and B.~Vicedo,
  ``On classical q-deformations of integrable sigma-models,''  
  JHEP {\bf 1311} (2013) 192  [arXiv:1308.3581 [hep-th]]. 

\bibitem{ORU}
  D.~Orlando, S.~Reffert and L.~I.~Uruchurtu,
  ``Classical integrability of the squashed three-sphere, warped AdS3 and
  Schr$\ddot{\rm o}$dinger spacetime via T-Duality,''
  J.\ Phys.\ A  {\bf 44} (2011) 115401.
  [arXiv:1011.1771 [hep-th]]. 

\bibitem{BR}
    B.~Basso and A.~Rej,
   ``On the integrability of two-dimensional models with $U(1) \times SU(N)$ symmetry,''  
    Nucl.\ Phys.\ B {\bf 866} (2013) 337  [arXiv:1207.0413 [hep-th]]. 
  
\bibitem{KY-Sch}
  I.~Kawaguchi and K.~Yoshida,
  ``Classical integrability of Schr\"odinger sigma models and $q$-deformed Poincare symmetry,''  
JHEP {\bf 1111} (2011) 094  [arXiv:1109.0872 [hep-th]]; 
  ``Exotic symmetry and monodromy equivalence in Schr\"odinger sigma models,''  
JHEP {\bf 1302} (2013) 024  [arXiv:1209.4147 [hep-th]]. \\ 
  I.~Kawaguchi, T.~Matsumoto and K.~Yoshida,
  ``Schr\"odinger sigma models and Jordanian twists,''  
JHEP {\bf 1308} (2013) 013  [arXiv:1305.6556 [hep-th]].  

\bibitem{Kame}
   T.~Kameyama and K.~Yoshida,
   ``Anisotropic Landau-Lifshitz sigma models from $q$-deformed AdS$_5 \times$S$^5$ superstrings,''
    JHEP {\bf 1408} (2014) 110
  [arXiv:1405.4467 [hep-th]]; 
   ``String theories on warped AdS backgrounds and integrable deformations of spin chains,''  
   JHEP {\bf 1305} (2013) 146  [arXiv:1304.1286 [hep-th]]. 

\bibitem{KOY}
  I.~Kawaguchi, D.~Orlando and K.~Yoshida,
  ``Yangian symmetry in deformed WZNW models on squashed spheres,''
  Phys.\ Lett.\  B {\bf 701} (2011) 475. 
  [arXiv:1104.0738 [hep-th]]; 
  I.~Kawaguchi and K.~Yoshida,
  ``A deformation of quantum affine algebra in squashed WZNW models,'' 
  J.\ Math.\ Phys.\  {\bf 55} (2014) 062302
  [arXiv:1311.4696 [hep-th]]. 

\bibitem{DMV-WZW} 
 F.~Delduc, M.~Magro and B.~Vicedo,
  ``Integrable double deformation of the principal chiral model,''
  Nucl.\ Phys.\ B {\bf 891} (2015) 312
  [arXiv:1410.8066 [hep-th]]. 
  
\bibitem{S-lambda}
  K.~Sfetsos,
  ``Integrable interpolations: From exact CFTs to non-Abelian T-duals,''
  Nucl.\ Phys.\ B {\bf 880} (2014) 225
  [arXiv:1312.4560 [hep-th]].  

\bibitem{MY-YBE}
 T.~Matsumoto and K.~Yoshida,
  ``Yang-Baxter sigma models based on the CYBE,''
  Nucl.\ Phys.\ B {\bf 893} (2015) 287
  [arXiv:1501.03665 [hep-th]]. 


\bibitem{Sch3}
  D.~Israel, C.~Kounnas, D.~Orlando and P.~M.~Petropoulos,
  ``Electric/magnetic deformations of S$^3$ and AdS$_3$, and geometric cosets,''
  Fortsch.\ Phys.\  {\bf 53} (2005) 73
  [hep-th/0405213].
    
\bibitem{DMV2}
  F.~Delduc, M.~Magro and B.~Vicedo,
  ``An integrable deformation of the AdS$_5\times$S$^5$ superstring action,''  
 Phys.\ Rev.\ Lett.\  {\bf 112} (2014) 051601
  [arXiv:1309.5850 [hep-th]];   
  ``Derivation of the action and symmetries of the $q$-deformed AdS$_5\times$S$^5$ superstring,'' 
  JHEP {\bf 1410} (2014) 132
  [arXiv:1406.6286 [hep-th]].  
   
  
\bibitem{KMY-Jordanian-typeIIB}
  I.~Kawaguchi, T.~Matsumoto and K.~Yoshida,
  ``Jordanian deformations of the AdS$_5\times$S$^5$ superstring,''
  JHEP {\bf 1404} (2014) 153
  [arXiv:1401.4855 [hep-th]].

\bibitem{LM}
 O.~Lunin and J.~M.~Maldacena,
  ``Deforming field theories with $U(1) \times U(1)$ global symmetry and their gravity duals,''  
JHEP {\bf 0505} (2005) 033  [hep-th/0502086].

\bibitem{Frolov}
   S.~Frolov,
   ``Lax pair for strings in Lunin-Maldacena background,''
   JHEP {\bf 0505} (2005) 069
   [hep-th/0503201].
  
\bibitem{HI}
  A.~Hashimoto and N.~Itzhaki,
  ``Noncommutative Yang-Mills and the AdS / CFT correspondence,''
  Phys.\ Lett.\ B {\bf 465} (1999) 142
  [hep-th/9907166].

\bibitem{MR}
  J.~M.~Maldacena and J.~G.~Russo,
  ``Large N limit of noncommutative gauge theories,''
  JHEP {\bf 9909} (1999) 025
  [hep-th/9908134].  



\bibitem{MMT}
C.~P.~Herzog, M.~Rangamani and S.~F.~Ross,
  ``Heating up Galilean holography,''
  JHEP {\bf 0811} (2008) 080
  [arXiv:0807.1099 [hep-th]]; 
 J.~Maldacena, D.~Martelli and Y.~Tachikawa,
  ``Comments on string theory backgrounds with non-relativistic conformal symmetry,''
  JHEP {\bf 0810} (2008) 072
  [arXiv:0807.1100 [hep-th]]; 
A.~Adams, K.~Balasubramanian and J.~McGreevy,
  ``Hot Spacetimes for Cold Atoms,''
  JHEP {\bf 0811} (2008) 059
  [arXiv:0807.1111 [hep-th]]. 
 
\bibitem{BK}
   N.~Bobev and A.~Kundu,
   ``Deformations of Holographic Duals to Non-Relativistic CFTs,''
   JHEP {\bf 0907} (2009) 098
   [arXiv:0904.2873 [hep-th]].

 \bibitem{BKP}
   N.~Bobev, A.~Kundu and K.~Pilch,
   ``Supersymmetric IIB Solutions with Schr\"odinger Symmetry,''
   JHEP {\bf 0907} (2009) 107
   [arXiv:0905.0673 [hep-th]].  

\bibitem{dipole1}  
 A.~Bergman and O.~J.~Ganor,
  ``Dipoles, twists and noncommutative gauge theory,''
  JHEP {\bf 0010} (2000) 018
  [hep-th/0008030].        

\bibitem{dipole2}  
 K.~Dasgupta, O.~J.~Ganor and G.~Rajesh,
  ``Vector deformations of N=4 superYang-Mills theory, pinned branes, and arched strings,''
  JHEP {\bf 0104} (2001) 034
  [hep-th/0010072].  

\bibitem{dipole3} 
  A.~Bergman, K.~Dasgupta, O.~J.~Ganor, J.~L.~Karczmarek and G.~Rajesh,
  ``Nonlocal field theories and their gravity duals,''
  Phys.\ Rev.\ D {\bf 65} (2002) 066005
  [hep-th/0103090].
    
\bibitem{dipole4}  
  M.~Alishahiha and O.~J.~Ganor,
  ``Twisted backgrounds, PP waves and nonlocal field theories,''
  JHEP {\bf 0303} (2003) 006
  [hep-th/0301080].

\bibitem{Imeroni}
  E.~Imeroni,
  ``On deformed gauge theories and their string/M-theory duals,''
  JHEP {\bf 0810} (2008) 026
  [arXiv:0808.1271 [hep-th]].
   
\bibitem{LM-MY}
  T.~Matsumoto and K.~Yoshida,
  ``Lunin-Maldacena backgrounds from the classical Yang-Baxter equation 
- towards the gravity/CYBE correspondence,''
  JHEP {\bf 1406} (2014) 135
  [arXiv:1404.1838 [hep-th]].    

\bibitem{MR-MY}  
 T.~Matsumoto and K.~Yoshida,
  ``Integrability of classical strings dual for noncommutative gauge theories,''
  JHEP {\bf 1406} (2014) 163 
  [arXiv:1404.3657 [hep-th]].  
  
\bibitem{Sch-MY}
  T.~Matsumoto and K.~Yoshida,
  ``Schr\"odinger geometries arising from Yang-Baxter deformations,'' 
  JHEP {\bf 1504} (2015) 180 [arXiv:1502.00740 [hep-th]].  

\bibitem{SUGRA-KMY}
 I.~Kawaguchi, T.~Matsumoto and K.~Yoshida,
  ``A Jordanian deformation of AdS space in type IIB supergravity,'' 
  JHEP {\bf 1406} (2014) 146 [arXiv:1402.6147 [hep-th]].    

\bibitem{MY-duality}
  T.~Matsumoto and K.~Yoshida,
  ``Yang-Baxter deformations and string dualities,''
  JHEP {\bf 1503} (2015) 137 [arXiv:1412.3658 [hep-th]]. 

\bibitem{Stijn}
S.~J.~van Tongeren,
  ``On classical Yang-Baxter based deformations of the AdS$_5 \times$S$^5$ superstring,''
  arXiv:1504.05516 [hep-th].

\bibitem{Benoit}
B.~Vicedo,
  ``Deformed integrable $\sigma$-models, classical $R$-matrices 
and classical exchange algebra on Drinfel'd doubles,''
  arXiv:1504.06303 [hep-th].
    
\bibitem{BZ}
P.~Basu and L.~A.~Pando Zayas,
  ``Chaos Rules out Integrability of Strings in AdS$_5 \times T^{1,1}$,''  
Phys.\ Lett.\ B {\bf 700} (2011) 243  [arXiv:1103.4107 [hep-th]].  

\bibitem{CO}
  A.~Catal-Ozer,
  ``Lunin-Maldacena deformations with three parameters,''
  JHEP {\bf 0602} (2006) 026
  [hep-th/0512290].

\bibitem{CMY}
  P.~M.~Crichigno, T.~Matsumoto and K.~Yoshida,
  ``Deformations of $T^{1,1}$ as Yang-Baxter sigma models,''
   JHEP {\bf 1412} (2014) 085
  [arXiv:1406.2249 [hep-th]]. 

\bibitem{MY-summary}
  T.~Matsumoto and K.~Yoshida,
  ``Integrable deformations of the AdS$_{5} \times S^5$ superstring 
and the classical Yang-Baxter equation 
{\it -Towards the gravity/CYBE correspondence-},''
  J.\ Phys.\ Conf.\ Ser.\  {\bf 563} (2014) 1,  012020
  [arXiv:1410.0575 [hep-th]].

\bibitem{Melvin}
M.~A.~Melvin, ``Pure Magnetic And Electric Geons,'' Phys.\ Lett.\ {\bf 8} (1964) 65. 

\bibitem{NW}
C.~R.~Nappi and E.~Witten,
  ``A WZW model based on a nonsemisimple group,''
  Phys.\ Rev.\ Lett.\  {\bf 71} (1993) 3751
  [hep-th/9310112].

\bibitem{SYY}
  S.~Schafer-Nameki, M.~Yamazaki and K.~Yoshida,
  ``Coset Construction for Duals of Non-relativistic CFTs,''
  JHEP {\bf 0905} (2009) 038
  [arXiv:0903.4245 [hep-th]].         

\bibitem{Gibbons:1987ps}
  G.~W.~Gibbons and K.~i.~Maeda,
  ``Black Holes and Membranes in Higher Dimensional Theories with Dilaton Fields,''
  Nucl.\ Phys.\ B {\bf 298} (1988) 741.

\bibitem{Tseytlin:1994ei}
  A.~A.~Tseytlin,
  ``Melvin solution in string theory,''
  Phys.\ Lett.\ B {\bf 346} (1995) 55
  [hep-th/9411198].

\bibitem{HT}
 A.~Hashimoto and K.~Thomas,
  ``Dualities, twists, and gauge theories with non-constant non-commutativity,''
  JHEP {\bf 0501} (2005) 033
  [hep-th/0410123].

\bibitem{HS}
 A.~Hashimoto and S.~Sethi,
  ``Holography and string dynamics in time dependent backgrounds,''
  Phys.\ Rev.\ Lett.\  {\bf 89} (2002) 261601
  [hep-th/0208126].

\bibitem{STV}
 M.~Spradlin, T.~Takayanagi and A.~Volovich,
  ``String theory in beta deformed spacetimes,''
  JHEP {\bf 0511} (2005) 039
  [hep-th/0509036].

\bibitem{Tai}
T.~S.~Tai
  ``D-branes in Lorentzian Melvin Geometry,''
  Prog.\ Theor.\ Phys.\ {\bf 117} (2007) 347
   [hep-th/0601039].

\bibitem{Drinfeld1}
  V.~G.~Drinfel'd,
  ``Hopf algebras and the quantum Yang-Baxter equation,'' 
  Sov.\ Math.\ Dokl.\ {\bf 32} (1985) 254.  

\bibitem{Drinfeld2}
 V.~G.~Drinfel'd,
  ``Quantum groups,''
  J.\ Sov.\ Math.\  {\bf 41} (1988) 898 
  [Zap.\ Nauchn.\ Semin.\  {\bf 155}, 18 (1986)].
  
  \bibitem{Jimbo}
  M.~Jimbo,
  ``A $q$ difference analog of $U(g)$ and the Yang-Baxter equation,''
  Lett.\ Math.\ Phys.\  {\bf 10} (1985) 63.        
  
\bibitem{Ricci:2007eq}
  R.~Ricci, A.~A.~Tseytlin and M.~Wolf,
  ``On T-Duality and Integrability for Strings on AdS Backgrounds,''
  JHEP {\bf 0712} (2007) 082
  [arXiv:0711.0707 [hep-th]].

\bibitem{Mohammedi:2008vd}
  N.~Mohammedi,
  ``On the geometry of classically integrable two-dimensional non-linear sigma models,''
  Nucl.\ Phys.\ B {\bf 839} (2010) 420
  [arXiv:0806.0550 [hep-th]].

\bibitem{Curtright:1994be}
  T.~Curtright and C.~K.~Zachos,
  ``Currents, charges, and canonical structure of pseudodual chiral models,''
  Phys.\ Rev.\ D {\bf 49} (1994) 5408
  [hep-th/9401006].

\bibitem{Beisert:2008iq}
  N.~Beisert, R.~Ricci, A.~A.~Tseytlin and M.~Wolf,
  ``Dual Superconformal Symmetry from AdS$_5\times$S$^5$ Superstring Integrability,''
  Phys.\ Rev.\ D {\bf 78} (2008) 126004
  [arXiv:0807.3228 [hep-th]].
  
\bibitem{Orlando:2012hu} 
  D.~Orlando and L.~I.~Uruchurtu,
  ``Integrable Superstrings on the Squashed Three-sphere,''
  JHEP {\bf 1210}, 007 (2012)
  [arXiv:1208.3680 [hep-th]].

\bibitem{LRT}
 O.~Lunin, R.~Roiban and A.~A.~Tseytlin,
  ``Supergravity backgrounds for deformations of AdS$_{n} \times$S$^n$ supercoset string models,''
  Nucl.\ Phys.\ B {\bf 891} (2015) 106
  [arXiv:1411.1066 [hep-th]].

\bibitem{SW}
  N.~Seiberg and E.~Witten,
  ``String theory and noncommutative geometry,''
  JHEP {\bf 9909} (1999) 032
  [hep-th/9908142].

\bibitem{NCOS}
R.~Gopakumar, S.~Minwalla, N.~Seiberg and A.~Strominger,
  ``(OM) theory in diverse dimensions,''
  JHEP {\bf 0008} (2000) 008
  [hep-th/0006062].

\bibitem{AGM} 
O.~Aharony, J.~Gomis and T.~Mehen,
  ``On theories with lightlike noncommutativity,''
  JHEP {\bf 0009} (2000) 023
  [hep-th/0006236].

\bibitem{Hull} 
 C.~M.~Hull,
  ``Timelike T duality, de Sitter space, large N gauge theories and topological field theory,''
  JHEP {\bf 9807} (1998) 021
  [hep-th/9806146]. 
  
\bibitem{ABF}
    G.~Arutyunov, R.~Borsato and S.~Frolov,
  ``S-matrix for strings on $\eta$-deformed AdS$_5\times$S$^5$,'' 
  JHEP {\bf 1404} (2014) 002 [arXiv:1312.3542 [hep-th]].

\bibitem{HOR}
  S.~Hellerman, D.~Orlando and S.~Reffert,
  ``String theory of the Omega deformation,''
  JHEP {\bf 1201} (2012) 148
  [arXiv:1106.0279 [hep-th]]; 
  ``The Omega Deformation From String and M-Theory,''
  JHEP {\bf 1207} (2012) 061
  [arXiv:1204.4192 [hep-th]]. \\
D.~Orlando and S.~Reffert,
  ``Deformed supersymmetric gauge theories from the fluxtrap background,''
  Int.\ J.\ Mod.\ Phys.\ A {\bf 28} (2013) 1330044
  [arXiv:1309.7350 [hep-th]]. \\ 
 N.~Lambert, D.~Orlando and S.~Reffert,
  ``Alpha- and Omega-Deformations from fluxes in M-Theory,''
  JHEP {\bf 1411} (2014) 162
  [arXiv:1409.1219 [hep-th]].
 

\end{thebibliography}
\end{document}